\documentclass[showpacs,nofootinbib]{revtex4}

\usepackage{amsfonts}
\usepackage{amsbsy} 
\usepackage{epsfig}
\usepackage{latexsym}
\input amssym.def
\input amssym.tex

\def\ben{\begin{equation}}
\def\een{\end{equation}}
\def\bea{\begin{eqnarray}}

\def\eea{\end{eqnarray}}

\advance \topmargin by -\headheight
\advance \topmargin by -\headsep
\evensidemargin \oddsidemargin
\advance\hoffset by -3mm  
\advance\voffset by  8mm  

\def\bR{\Bbb R}

\def\b1{e^0}

\begin{document}
\title{Disappearance and emergence of space and time in quantum gravity}

\begin{flushright}AEI-2013-050\end{flushright}

\author{{\bf Daniele Oriti}\footnote{daniele.oriti@aei.mpg.de}}
\affiliation{Max Planck Institute for Gravitational Physics (Albert Einstein Institute) \\ Am M\"uhlenberg 1, D-14476 Golm, Germany, EU}
\begin{abstract}
We discuss the hints for the disappearance of continuum space and time at microscopic scale. These include arguments for a discrete nature of them or for a fundamental non-locality, in a quantum theory of gravity. We discuss how these ideas are realized in specific quantum gravity approaches. Turning then the problem around, we consider the emergence of continuum space and time from the collective behaviour of discrete, pre-geometric atoms of quantum space, and for understanding spacetime as a kind of "condensate", and we present the case for this emergence process being the result of a phase transition, dubbed \lq\lq geometrogenesis\rq\rq. We discuss some conceptual issues of this scenario and of the idea of emergent spacetime in general. As a concrete example, we outline the GFT framework for quantum gravity, and illustrate a tentative procedure for the emergence of spacetime in this framework. Last, we re-examine the conceptual issues raised by the  emergent spacetime scenario in light of this concrete example.
\end{abstract}


\maketitle
\section{Introduction: philosophy of quantum gravity \lq\lq in the making\rq\rq}
Quantum Gravity is a hard problem. Beside the technical difficulties of developing and applying the right mathematics, adapted to a background independent context, the main difficulties are conceptual. One is to imagine first, and then identify in mathematical terms, the fundamental degrees of freedom of the theory. The next one is to show how the quantum dynamics of such fundamental degrees of freedom leads to an effective classical behaviour described, at least in some approximation, by General Relativity. If the fundamental degrees of freedom are not spatio-temporal in the usual sense, that is, they  are not associated to a smooth spacetime manifold (as the basic variable of GR, the spacetime metric), then one faces an additional challenge: to show that the spacetime continuum itself, and the relative geometric variables\footnote{In this paper, we focus on pure gravity and geometry, leaving aside the coupling (or the emergence) of matter and gauge fields. While this is of course an important issue, we do not feel that the problem of emergence of spacetime, as phrased here will be much affected by adding other degrees of freedom in the picture.} arise at least as a useful approximation to the more fundamental structures the theory is based on. 

There has been an enormous progress in several approaches to quantum gravity, in recent years, and in particular in the ones we will be focusing on, group field theories, spin foam models and loop quantum gravity. This has led to concrete suggestions  for the fundamental degrees of freedom, to interesting proposals for the fundamental quantum dynamics, and even physical applications in simplified context (e.g symmetry reduced models in cosmology). The \lq\lq problem of the continuum\rq\rq or \lq\lq the issue of the emergence of spacetime and geometry\rq\rq remains open. This is , in our opinion, the most pressing issue that quantum gravity approaches have to tackle. Indeed, it is the real challenge even assuming that the true fundamental degrees of freedom of quantum gravity have been found. Such degrees of freedom would be of a different nature than anything else we have learned to deal with up to now, in usual spacetime-based physics, and many of the tools that we have developed to describe similar micro-to-macro and few-to-many transitions in ordinary physical systems, e.g. in condensed matter theory, would have to be at the very least adapted to this new background independent context and reinterpreted accordingly. Beyond the technical difficulties, this is necessarily going to be a conceptual tour-de-force. 

\

It is here that philosophy can and should help the development of theoretical physics. Clarify the conceptual basis of quantum gravity approaches, smoothen the tortuous reasoning of theoretical physicists towards a solution of the continuum conundrum, alert them of hidden assumptions or prejudices and of shaky conceptual foundations for their theories, force them to be honest and rigorous in their thinking, and not only in their mathematics. This is what we believe philosophers working on the foundations of spacetime theories, and of quantum gravity in particular, should do. In this respect, we cannot but agree with Huggett and W\"utrich  \cite{H-W} on the need to  perform philosophical investigations {\it in parallel} with the formal development of the theory, rather than arrive on the stage after the show has taken place, only to elaborate on it, while theoretical physicists are then struggling with preparing a new one\footnote{We believe that philosophical analysis should take place in parallel with scientific developments also when its object is the scientific process itself, its method, its sociological aspects, its modes of development, i.e. philosophy of science proper and epistemology. The risk, otherwise, is to do not an ethology of scientists, but their paleontology.}. 

The obvious problem is that a philosophical reflection based on the ongoing, tentative work of theoretical physicists on models that, most likely, will turn out to be incorrect or only partially understood in the future will also be resting on shaky grounds, sharing the same risks of future irrelevance. In addition, if the ground on which these philosophical reflections take place is that of quantum gravity theories, they will also have to struggle among very abstract or yet-to-be developed mathematics and some conceptual darkness, and without much input from experiments (which is much less true in other areas of theoretical physics).  Well, the only reaction we feel is appropriate to these last worries is: \lq\lq Welcome to the club!\rq\rq 

Another point we agree on with Huggett and W\"utrich \cite{H-W} is that philosophical reflection, if it has to be serious and concrete, needs concrete examples, on which general reasoning can be applied. 

Thus, in this paper, we offer to philosophers a brief discussion of the hints of a non-fundamental nature of  continuum space and time and the usual spacetime physics, a flying survey of how these hints are realized in some quantum gravity approaches, a scenario for the emergence of spacetime in quantum gravity, a concrete framework for realizing this scenario (the group field theory framework) and an explicit (if tentative) working example of what such emergence process could be like, in this framework.
Along the way, while we will not enter into any detailed discussion of philosophical issues related to the idea of emergent spacetime (let alone emergence tout court), we will not refrain for commenting on some of them. 

Our treatment will be necessarily rather sketchy and utterly incomplete, and we will try to refer to the relevant literature whenever needed. Still, we hope that we will manage to provide philosophers and physicists alike some useful and possibly stimulating material, for exercising philosophical reflections about the foundations (and emergence) of space and time.

\section{Hints for the disappearance of space and time}
The current description of the world is as a collection of (quantum) fields living on a spacetime continuum manifold, and interacting locally (which is strictly related to the use of continuum manifold as a model of spacetime). In particular, the structure and physical properties spacetime itself are encoded in one such field, the metric, and the most conservative option is to apply standard quantization methods to the dynamics of the metric field. There are several reasons why this strategy either did not work or proved very difficult. But the hints for the disappearance of spacetime can be taken as suggestions that it is the very starting point of this strategy has to be replaced by something more radical\footnote{As we will discuss, the most successful example of such conservative strategy, canonical loop quantum gravity \cite{LQG}, ends up anyway replacing the spacetime continuum with something more radical. In fact, the idea of emergent space and time and the non-existence of space and time at the fundamental level have been repeatedly advocated by Rovelli \cite{carlo}, focusing on the temporal aspects, even though from a different perspective.}. We now recall some of them, briefly.

\

One preliminary thing to notice is that, even if the strategy of simply applying standard quantization methods to the metric field was successful (and within the usual interpretative framework of quantum mechanics now applied to spacetime as a whole), the resulting picture of spacetime would be highly non-standard. The basic quantum observables would be operators corresponding to the metric field (or to equivalent geometric quantities, like the gravitational connection) and quantum states would carry a representation of such observables. Straightforward consequences of the quantum formalism would be the quantum uncertainties in the determination of the same geometric quantities, possibly a discrete structure of their spectra, the superposition of metric states and thus of causal structures, and so on. A description of spacetime in which these phenomena take place could hardly be considered a minor modification of the classical one. Still, one could argue that the resulting weirdness would be purely a result of the quantization process, to be removed in a classical limit. In a sense, the usual continuum spacetime concepts would simply be mapped to weirder ones, their quantum counterparts, by the quantization map, but not dismissed altogether for a new conceptual scheme. As said, to cope conceptually with the resulting quantum framework would be already challenging (as it can be seen in simplified schemes like quantum cosmologies \cite{LQC}), but something even more radical seems to be needed\footnote{Notice that we are also neglecting here the conceptual implications of classical and quantum diffeomorphism invariance in the continuum, e.g. the infamous \lq\lq problem of time\rq\rq \cite{time}.}. 

\

The notion of {\it spacetime localization} is challenged already in semi-classical GR, i.e. simply considering quantum fields in a curved spacetime, the intuitive idea being that any attempt at exact localization would imply probing spacetime events with such a concentration of energy in such small spacetime regions, that the event itself would be hidden by the formation of an horizon. This idea is at the root of some models of quantum spacetime based on non-commutative geometry \cite{doplicher}, and it is only one of the many general arguments for the existence of a minimal length scale, usually associated to the Planck scale, beyond which the local description of events of a spacetime continuum ceases to make sense. The associated model building, aimed at an effective level of description of more fundamental quantum gravity structures, is vast and the basis for much of current quantum gravity phenomenology (see \cite{sabine} for a review). Non-locality seems an inevitable feature of most models based on some form of spacetime non-commutativity, where the existence of a minimal length translates into non-vanishing commutation relation for (operators corresponding to) spacetime coordinates. One model that takes this feature seriously is so-called {\it relative locality} \cite{RelatLoc}, a modern incarnation of the idea of {\it deformed or doubly special relativity} \cite{DSR}, in turn a specific version of the effective models of quantum spacetime based on non-commutative geometry and a minimal length. Here the starting point is the idea that what we really measure, in physical experiments, are momenta (and energies), while spacetime has to be reconstructed our of such momentum measurement, plus the idea that a full theory of quantum gravity should allow for a non-trivial geometry of momentum space itself, as well as a curved geometry of spacetime as in GR. A generic consequence of a curved momentum space seems in fact to be the relativity of the notion of locality of processes in spacetime, and thus a breakdown of the universality of reconstruction of a spacetime manifold for physical events, for different observers.
Similar arguments and supporting partial results have been put forward in string theory \cite{strings}, where the very fact that spacetime geometry is on the one hand probed and on the other hand constituted by extended objects leads to a non-local description of it, allowing for example for several dual descriptions (the AdS/CFT conjectured duality is also often presented as evidence for an emergent nature of spacetime) \cite{stringdualities, seiberg}.

\

Spacetime singularities in GR, the cosmological singularity at the Big Bang and black holes singularities, are generally regarded as a most convincing hint that a quantum theory of gravity is needed. They can be given a minimalistic interpretation, akin to singularities in the classical electromagnetic field at the center of a spherically symmetric source (e.g. in the naive classical model of the hydrogen atom). Then, they would simply suggest the need for quantum corrections to the dynamics of geometry and matter \cite{LQC}. More radically, they have been argued to be more than a sign of incompleteness of the classical description of spacetime, and rather of breakdown of the very notion of spacetime continuum, itself implying the need for new degrees of freedom and a change of descriptive framework. This second interpretation is more in line with the prominent role that divergences and singularities in physical quantities play in quantum and statistical field theory, as a signal of a change of the effective dynamics with scale due to new degrees of freedom becoming relevant, to be taken care of via the renormalization group, or as a signal of a phase transition of a physical system, due to the collective dynamics of its microscopic constituents (see the discussion on the importance of singularities in relation to the issue of emergent physics \cite{battermann, butterfield}). We will return to the issue of cosmological singularities, from this perspective, in the following. The analysis of cosmological singularities in GR and supergravity theories, in particular via the so-called BKL approach, has led to another argument for the disappearance of spacetime in such extreme situations. The result of such analysis \cite{nicolai} is that, at least in some approximation, the dynamics of spacetime (and matter) close to the singularity can be mapped to the dynamics of a point particle on an infinite-dimensional coset space based on the Kac-Moody algebra $E_{10}$. In turn, this suggest that a new, purely algebraic description of the relevant physics becomes appropriate at and beyond the singularity, one that would not be based on any notion of spacetime at all. 

\

In the case of black holes, the attention has focused on the physics at the horizon and at the associated thermodynamics. It has been argued \cite{giddings} that the black hole evolution can be unitary and thus information preserving, despite their evaporation, only at the expense of locality.  But it is the notion of black hole entropy (or more generally of horizon entropy \cite{tedrenaud}) itself that challenges in a very serious way the usual continuum picture of spacetime. A black hole horizon is nothing but a particular region of spacetime, so it is spacetime itself that has entropy. And something that has entropy has a microstructure, whose number of degrees of freedom that entropy counts. If spacetime was a continuum, that entropy would be infinite. So the finite value for the horizon entropy is a clear indication that there exist a discrete microstructure for spacetime \cite{sorkinBH}. The laws of black hole thermodynamics hint at something even deeper. In \cite{jacobson} Jacobson showed that the relation between Einstein's GR equations and the laws of horizon thermodynamics can be turned upside down, in the following sense\footnote{We hope the more sophisticated reader will forgive us for the unsophisticated nature of our account of this topic.}. Considering a small region of an arbitrary spacetime, the metric $g$ can be approximated locally by a flat one and a local Rindler horizon can be identified (this is the horizon associated to uniformly accelerated motion in the region). The Rindler horizon isolates the subregion accessible to uniformly accelerated observers and can itself be shown to have a thermal character, i.e. to have a temperature $T$ and an entropy $S$, via entanglement arguments which also show that its entropy is proportional to its surface area (thus a function of the metric)\footnote{The proportionality factor depends on the precise form of the Equivalence Principle that is assumed to hold.}. Assuming that there are also matter fields $\phi$ around, one can compute their energy flux $Q$ across the same horizon (from their energy-momentum tensor, itself a function of the metric). Using another few reasonable assumptions (e.g. diffeomorphism invariance and energy conservation), one can then show that the first law of thermodynamics, i.e. the equation of state $T\delta S = \delta Q$, is equivalent to the Einstein's dynamical equations for the metric and the matter fields. In other, more suggestive words, the Einstein's equations are but an equation of state for whatever microscopic degrees of freedom can be collectively described by a metric and some matter fields, and of which the functions $S(g)$ and $Q(g,\phi)$ define the entropy and the heat, in a macroscopic approximation where the system can be approximated by a spacetime manifold. These results have later been extended to cover more general gravitational theories and non-equilibrium situations.

\

To conclude our brief survey of hints at the disappearance of spacetime and geometry from the fundamental ontology of the world, let us mention the insights obtained from condensed matter analogues of gravitational phenomena \cite{analog, volovik}. The insights here are, at the moment, only at the kinematical level (but see \cite{sindoni}). However, they support further the idea that spacetime itself is an emergent concept and that metric and matter fields are only collective variables for more fundamental degrees of freedom. The basic analogy is captured by the following general fact (taken verbatim from \cite{analog}). If a fluid is barotropic and inviscid, and the flow is irrotational (though possibly time dependent) then the equation of motion for the velocity potential describing an acoustic disturbance is identical to the dÕAlembertian equation of motion for a minimally-coupled massless scalar field propagating in a (3+1)-dimensional Lorentzian geometry. Thus the sound wave couples not to the metric of the laboratory in which the whole fluids sits, but to an effective Lorentzian metric, which depends algebraically on the density, velocity of flow, and local sound speed in the fluid, i.e. by the hydrodynamic variables. When the fluid is non-homogeneous and flowing, the acoustic Riemann tensor associated with this Lorentzian metric will be nonzero. This fact can be further generalized, specific fluids (the most interesting being quantum fluids, like Bose condensates) can be studied and more analogues of gravitational phenomena can be identified. Clearly, both the effective metric and the effective scalar field (the sound waves) are only emergent, collective excitations of the underlying atomic system. The main limitation is dynamical. The emergent metric does not satisfy general relativistic dynamical equations, but the equations of hydrodynamics (even though some limited form of gravitational dynamics can be reproduced \cite{sindoni}). For now, this serves only as an inspiration for the idea that spacetime itself could be the result of a macroscopic approximation, valid in a particular phase of a more fundamental non-spatio-temporal system, for which continuum metric and matter fields are not appropriate variables.

\section{Microstructure of spacetime in various QG approaches}
\subsection{Brief survey}
Now that we have surveyed different reasons for believing that continuum spacetime may not be the fundamental substratum of a quantum gravity theory, let us survey, again briefly, the suggestions that different quantum gravity approaches offer as to what instead this substratum may be. 

Loop quantum gravity \cite{LQG} started off as a canonical quantization of continuum GR reformulated in terms of first order variables, namely a Lorentz connection and a triad field, defining the phase space of the theory. Still, the completion of the quantization programme, at least before the imposition of the dynamics (Hamiltonian constraint), leaves with a Hilbert space of the form $\mathcal{H}=\bigoplus_\gamma\mathcal{H}_\gamma $, i.e. a direct sum of Hilbert spaces associated to {\it graphs}. For each given graph $\gamma$, considered embedded in the spatial manifold where the canonical analysis takes place, a generic state would be given by a wave function $\psi_\gamma(g_1,....,g_E) $for an assignment of data to the elements of the graph, which are themselves a discrete version of the continuum variables: group elements $g_e$ representing holonomies of the gravitational connection along the links $e=1,...,E$ of the graph, or Lie algebra elements representing their conjugate discretized triad. Thus we have discrete combinatorial structures, the graphs, labelled by algebraic data only. The algebraic nature of the data is even more manifest in the fact that a complete orthonormal basis of such states is given by spin networks, in which the same graphs are labelled by representations of the Lorentz (or rotation) group. In later developments, the initial embedding of the graphs into a spatial manifold is dropped, so that one is really left with no reference to any underlying continuum spacetime, which has instead to be reconstructed from the combinatorial and algebraic data alone. The canonical dynamics is implemented by the action of an Hamiltonian constraint operator which acts by mapping any labelled graph into a linear superposition of graphs with different labels (in the general graph-changing case), thus retaining the purely combinatorial and algebraic character of the formalism. The same is true at the covariant level, where the dynamics is implemented by a sum over histories. Each history is defined by a 2-complex $\sigma$ (a collection of vertices, links and faces) labelled by the same algebraic data assigned to states. This is called a \lq spin foam\rq. A path integral-like dynamics (a \lq spin foam model\rq) is then specified by a probability amplitude weighting each spin foam, and by a sum over all possible algebraic data assigned to each 2-complex and, in general, by a sum over 2-complexes. In recent years, it has been realized that the graphs $\gamma$ are best understood as the 1-skeleton of a dual 3d cellular complex, usually taken to be a simplicial one (with consequent restriction on the valence of $\gamma$), with the Lie algebra elements labeling the links of $\gamma$ associated to the 2-cells of the cellular complex, and the 2-complexes $\sigma$ as the 2-skeleton of a 4d cellular (simplicial) complex. Also the choice of spin foam amplitudes is adapted to this simplicial setting, being usually given by or motivated from a lattice version of the gravitational path integral. The basic fact remains that the fundamental excitations of quantum spacetime are given by discrete and algebraic objects, rather than smooth quantum fields. It is natural, in this set-up, to interpret such excitations in {\it realistic} terms, and not as a mere regularization, devoid of physical significance; they are the theory's candidates for the fundamental quantum degrees of freedom of spacetime\footnote{Notice also that spin foam models can be defined (and in practice they usually are) independently of any canonical derivation, thus the spin foam complexes can have arbitrary topology and their algebraic labels (both in the bulk and in the boundary) can correspond to timelike data (edges, triangles etc).}.
 
This most recent incarnation of the LQG and spin foam approach is closer to other covariant simplicial gravity approaches, like quantum Regge calculus \cite{hamber} and (causal) dynamical triangulations \cite{CDT}.
Quantum Regge calculus is a straightforward lattice quantization of gravity. It defines a lattice path integral for the Regge action, a discretization of GR with metric variables represented by the lengths of the edges of the simplicial complex used to replace the spacetime manifold, and summed over to define a discrete sum over histories, with some appropriate choice of measure. (Causal) Dynamical triangulations are based on a sort of \lq complementary\rq definition of a discrete gravitational path integral: the edge lengths are fixed to be equal for every edge of the simplicial complex, while the sum over histories is defined by a sum over all such equilateral triangulations, usually restricted to a given (trivial) topology, weighted by the same (exponential of the) Regge action (restricted to the equilateral case).
No boundary Hilbert space is defined, as the focus is usually on the sum over histories definition of the theory, but boundary data are associated to 3d simplicial complexes, labelled by boundary edge lengths (in quantum Regge calculus), or assumed to be equal (in dynamical triangulations). 

The structures used are thus similar to the ones of LQG and spin foams. The main differences are the different types of variables labeling the combinatorial structures (group and Lie algebra elements or spins in the LQG case, interpreted as triangle areas, holonomies etc; real numbers interpreted as edge lengths in simplicial quantum gravity), and the choices of amplitudes for the quantum histories to be summed over, i.e. the precise definition of the quantum dynamics. All these approaches can be defined, at least in principle (i.e. modulo various technical complications), in both Riemannian and Lorentzian settings, the difference being encoded in the data assigned to the simplicial complex and corresponding modifications at the level of the quantum amplitudes. At a more conceptual level the main difference is that, while as mentioned the discrete excitations in LQG are naturally interpreted realistically, thus are given ontological significance, the same discrete structures in simplicial gravity are often interpreted as a regularization tool only, needed to give mathematical meaning to the theory which is, in the end, a continuum theory. This last attitude is the standard one, of course, in non-gravitational lattice gauge theory. This different interpretation has influenced the goals of the different research programmes. The main goal of the simplicial gravity approaches has been to prove, using statistical field theory methods, the existence of a phase transition of the discrete system in which any \lq\lq discretization artifact\rq\rq would be removed and that would then define \lq the continuum theory\rq. In LQG and spin foams, instead, the focus has been on extracting physics from the theory formulated in terms of the discrete spin networks and spin foams, and, due to obvious technical difficulties of doing otherwise, mainly in the regime of small number of degrees of freedom, i.e. simple graphs and simple cellular complexes.

Related (historically and mathematically) to the dynamical triangulations programme is the formalism of matrix models \cite{mm}, which provides in fact a successful quantization of 2d Riemannian gravity. The dynamics is given again by a sum over 2d triangulations, with no additional labels. The crucial difference is that this sum is obtained as the perturbative Feynman expansion of (the free energy of) a theory whose fundamental variables are abstract $N\times N$ matrices $M^i_{j}$ whose classical action has no direct relation to any classical continuum gravity dynamics, for example $S(M) = \frac{1}{2}tr(M^2) - \frac{\lambda}{N^{1/2}}tr(M^3)$. The matrices can be graphically represented as edges with endpoints labeled by their indices, and the Feynman diagrams of the theory are represented by triangulations, since the interaction vertex has the combinatorial structure of three edges glued along vertices to form a triangle, while the propagator corresponds to the identification of edges across two such triangles. 
At this discrete level one can only hope to an indirect connection with gravity and geometry analogous to that of dynamical triangulations. However, appropriate matrix models can be shown to undergo a phase transition, which can be phrased in terms of a \lq\lq condensation\rq\rq of the eigenvalues of the matrices (the true degrees of freedom) for large N and some critical value of the coupling constant $\lambda$, and to possess a continuum phase in which the theory they define matches (in computable quantum observables) 2d Liouville quantum gravity. While the formal setting is very close to the dynamical triangulations approach, the fact that the triangulations arise as Feynman diagrams of a theory of matrices suggests the possibility to give a more realistic interpretation to the same matrices as underlying pre-geometric degrees of freedom of 2d spacetime. It is in this sense that, for example, matrix models are presented as an instance of \lq\lq emergent spacetime\rq\rq in \cite{seiberg}.

Another example of discrete approach to quantum gravity is causal set theory \cite{CS}. Here the starting point is again the continuum theory but only in the observation that the metric of a given spacetime is entirely characterized by the causal structure, i.e. the set of causal relations between any pair of spacetime points, and the conformal factor or, equivalently, the volume element. Then, a discrete counterpart of the same data is postulated to be the fundamental substratum of the world: a discrete set of \lq\lq events\rq\rq , each assigned a constant volume element (usually assumed to be of the order (Planck length)$^4$), and their causal relations, i.e. a partially ordered set with a constant volume label on each element. The dynamics is seek once more in a covariant sum over histories setting, i.e. in a sum over causal sets weighted by some yet to be defined quantum amplitude (a quantum version of the recently developed classical sequential growth dynamics, which is a classical stochastic process, and involving the recently defined causal set discrete analogue of the Einstein action \cite{CS}). It is clear that the spirit of this approach is not to look for a regularization of a continuum theory, but to identify the true, more fundamental ontology of spacetime.

The type of discrete structures arising in spin foams and simplicial gravity can seem very different from causal sets. This is clearly true to some extent, but some similarities can be identified. More precisely, Lorentzian spin foam models can be constructed \cite{causalSF} in such a way that the 1-skeleton of the spin foam 2-complex is understood as a discrete set of events  (the vertices) with their causal relations, restricted to five per vertex (in four dimensions), i.e. to a sort of \lq\lq causal nearest neighbors\rq\rq, and the spin foam amplitudes can be defined so to reflect this underlying causal structure. Thus, the main kinematical differences between the approaches are a richer set of pre-geometric labels and a more restricted set of causal relations taken into account in spin foams with respect to causal set theory\footnote{The other difference is that the 1-skeleton of a spin foam complex is a directed set, rather than a poset, i.e. may contain closed timelike loops. These could be added also in standard causal set theory and it is unclear whether they would modify much the formalism.}. The last difference may result in a breaking (or deformation) of local Lorentz symmetry in a continuum approximation of spin foams, while causal sets are argued to be Lorentz invariant \cite{CS}, but on the other hand it allows for a discrete version of locality which facilitates the mathematical treatment of the theory (and which makes the group field theory formulation of it possible), while causal sets  are radically non-local.

\subsection{The GFT framework}
We give a few more details on the group field theory approach \cite{GFT}, since we will later give an explicit example of emergent spacetime in this context. 
The formalism is a field theory over a group manifold (or the corresponding Lie algebra) with the basic variable being a (complex) field $\phi(g_1,g_2,..,g_d)=\phi_{12..d}$ function of $d$ group elements, for models aiming at a quantization of a d-dimensional spacetime (the most relevant case, therefore, being $d=4$). It can be represented graphically as a (d-1)-simplex with field arguments associated to the faces of it, or as a d-valent graph vertex, with field arguments associated to the links. The dynamics is specified by a choice of action $S(\phi) = \int \phi \mathcal{K} \phi \,+\, \lambda\,\int \phi \phi... \phi \,\mathcal{V}$, characterized in particular by an interaction term with a non-standard (with respect to usual local QFT) convolution of fields in terms of their arguments (analogous to the tracing of indices in matrix models). The specific pattern of convolution chosen depends on which requirement replaces locality of standard QFTs. 
The quantum dynamics is given by the partition function, expanded in Feynman diagrams around the Fock vacuum:

$$
Z\,=\,\int\mathcal{D}\phi\,\,e^{i S(\phi)}\,=\,\sum_\Gamma\,A_\Gamma\qquad .
$$

The choice of action involves a choice of the kinetic term, and a choice of interaction. One example, in $d=4$ and for $g_i \in SO(4)$, is:
\begin{equation}
\label{actiongroup}
S = \frac{1}{2} \int [d g_i]^4 \phi^2_{1234}  + \frac{\lambda}{5!}  \int [d g_i]^{10}  
\phi_{1234} \,\phi_{4567} \,\phi_{7389}\, \phi_{962\,10} \,\phi_{10\,851} 
\end{equation}
where $d g$ is the normalized Haar measure, $\lambda$ is a coupling constant, and the field has the invariance $\phi(h g_1,..., h g_4) = \phi(g_1,...,g_4)$ for any $h\in SO(4)$. The combinatorics of convoluted field arguments in the interaction matches the combinatorics of five tetrahedra glued along common triangles to form a 4-simplex. The propagator induces a simple gluing of 4-simplices across one shared tetrahedron, when used in perturbative expansion to form Feynman diagrams. Such Feynman diagrams are representable, by construction, as 4d simplicial complexes. This model describes topological $SO(4)$ BF theory. Models aiming at the description of 4d gravity are constructed \cite{aristidedaniele, SF} imposing specific restrictions on the fields $\phi$, corresponding to discrete and algebraic versions of the constraints that reduce BF theory to gravity in four dimensions \cite{aristidedaniele, SF}. In these models, one could say that usual QFT locality is replaced by {\it simpliciality}, i.e. the requirement that interactions are associated to d-simplices.

Another class of models has been the focus of studies \cite{GFTrenorm} aiming at extending standard renormalization tools to GFTs and at proving renormalizability of specific models, relying heavily on the recent results of the simpler tensor models \cite{tensorReview}, and thus referred to as {\it tensorial GFTs} or TGFTs. In particular, TGFTs use a new notion of locality: \textit{tensor invariance} (related to invariance under $U(N)^d$ transformations, where $N$ is a cut-off on the Lie algebra dual to $G$). One labels the arguments of the basic field with an index $k$ and defines the convolutions such that any $k$-th index of a field $\phi$ is
contracted with a $k$-th index of a conjugate field $\overline{\phi}$. These invariant convolutions can be graphically represented as $d$-\textit{colored graphs}, constructed from two types of nodes and $d$ types of colored edges: each white (resp. black) dot
represents a field $\phi$ (resp. $\overline{\phi}$), while a contraction of two indices in position $k$ is associated to an edge with color label $k$. The interactions (with coupling constants $t_b$) are sum of connected invariants $I_b$
\begin{equation}
S(\phi , \overline{\phi}) = \sum_{b} t_b I_b (\phi , \overline{\phi})\,. 
\end{equation}
The kinetic term is taken to be the Laplace-Beltrami operator $\Delta$ acting on $G^{d}$, accompanied by a \lq\lq mass\rq\rq term.

\

There are two main ways of understanding GFTs. The first is as a second quantized field theory of spin network vertices, each corresponding to a {\it quantum} of the field $\phi$ and labelled by the $d$ group or Lie algebra elements, constructed in such a way that its quantum states are generic superpositions of spin networks and its Feynman diagrams are spin foams. For any given spin foam models, there exist a choice of GFT action (thus, a specific GFT model), such that the corresponding Feynman amplitudes are the chosen spin foam amplitudes. It can be seen, therefore, as a possible (second quantized) incarnation of the LQG programme. Using the dual simplicial formulation of spin networks and spin foams, in which each spin network vertex corresponds to a labelled simplex and each spin foam 2-complex is dual to a simplicial complex, GFTs can also be understood as a second quantization of simplicial geometry, in which simplices can be created and annihilated in fundamental interaction processes, as well as change size an shape (with group or Lie algebra elements characterizing areas of triangles, discrete curvature, etc). The combinatorics of field convolutions in the GFT interaction should correspond to the combinatorics of face gluings in a d-simplex. According to this simplicial interpretation and to the current treatment of spin foam models, the corresponding Feynman amplitudes are then related directly to discrete gravity. The group manifold $G$ chosen is the Lorentz group in the appropriate dimension or its Riemannian counterpart.

A second way is as a generalization of matrix models for $d=2$ gravity. This generalization takes place at two levels. The first is in dimension. Instead of matrices with two indices, represented as a segment with two endpoints, we have a field function of $d$ variables, represented as a (d-1)-simplex. Correspondingly, the Feynman diagrams of the theory are d-complexes, rather than 2d triangulations. The second level of generalization is in the data carried by the fields. Instead of finite index sets, the arguments of the field are now group manifolds or their Lie algebra (and one can go from one to the other by a generalized Fourier transform). These additional data are needed, formally, for bringing in the structures used in LQG and simplicial gravity, and, physically, because it can be expected that higher-dimensional gravity needs a much richer framework to be properly described at the quantum level, and these data can be understood as \lq\lq seeds\rq\rq of the geometry that has to emerge from the models. This of course brings mathematical complications, alongside conceptual and physical richness, and these models prove quite non-trivial\footnote{Much progress on the mathematical backbone of GFTs has been obtained in the simpler tensor models, in which the combinatorial structure is maintained but the extra degrees of freedom are dropped. Several insights have been obtained \cite{tensorReview}, most notably a well-defined notion of large-N limit, and can now be incorporated in GFTs.}.

The main advantages compared to other formulations of LQG or spin foams are first of all that GFTs offer a {\it complete} definition of the quantum dynamics, e.g. a clear prescription for the weights to be used in the sum over spin foams and for how this should be generated; second, that one can take advantage of more or less standard QFT tools in studying the theory, despite the fully background independent context (from the point of view of physical spacetime). This in particular would be a key asset to study the physics of large numbers of LQG degrees of freedom (large spin networks and spin foams), e.g. using powerful tools like the renormalization group \cite{vincent}. At the same time, it calls for a new perspective on the theory, in the sense that, holding again to a realistic interpretation of the discrete structures used in in LQG and spin foams, it suggests seeing the underlying field theory generating them as {\it the definition} of the dynamics, and the place to look for its physical consequences, rather than as an auxiliary tool. Also, it suggests that the fundamental dynamics, encoded in the GFT action, should not be necessarily derived from any canonical quantization of the continuum theory, but something simpler, intrinsically discrete, even if motivated by continuum considerations (this is the road already taken by spin foam models). Compared with other simplicial gravity approaches, similar structures are used as discrete counterparts of continuum spacetime (labelled triangulations) and the same covariant implementation of the dynamics via sum over histories. This dynamics looks like a combination of both quantum Regge calculus and dynamical triangulations, with a sum over triangulations each weighted by a sum over discrete geometric data (albeit the variables chosen are different), weighted by a quantum amplitude that indeed can be directly related to a simplicial gravity path integral. Also, as mentioned, similar methods of analysis are called for, taken from statistical field theory, and similar goals: look for continuum approximations/limits, explore the phase structure of the models. Beyond the technical differences, however, the main difference is probably conceptual, in that (borrowing from the LQG camp) GFTs call for a re-interpretation of the same structures in realistic terms, as physical (discrete and quantum) degrees of freedom of  spacetime. 

\subsection{Some more comments}
Before moving to the issue of emergence of spacetime, let us add some assorted comments on the discrete and quantum picture of spacetime portrayed by these approaches, and by GFT in particular.
A similar brief survey of QG approaches in relation to the issue of emergent spacetime was given very nicely in \cite{H-W}. Compared to that account, beside the addition of the GFT formalism to the picture, the main comment we have is that in our opinion the distinction between these various approaches is more blurred than portrayed there (as the GFT approach shows explicitly). Simplicial lattices can be used in correspondence with spin network graphs and spin foams. Elements of these combinatorial structures can be understood in causal terms, as done in causal sets. Quantum superposition of discrete structures is a key ingredient of all these approaches. Most of them involve a sum over topologies as well (this is inevitable in GFTs, in particular). 

However, it is true, as mentioned, that this merging of structures and methods may call for a reinterpretation of them.  
For one thing, the lattices used in LQG and spin foams and GFT are interpreted more realistically than in simplicial quantum gravity approaches. At the same time, the recasting of spin foams as Feynman diagrams of a GFT suggest that, while they are indeed possible physical processes of quantum building blocks of spacetime, the physics they encode is to be extracted from their sum and not from them individually, if not in very peculiar approximations and after renormalization. 

The other main point is the extent to which these discrete structures should be interpreted in spatio-temporal terms. Here we refer to both the combinatorial structures, graphs and cellular complexes, and to the labels assigned to them, usually interpreted in terms of discrete geometry. In our opinion, the spatio-temporal meaning of these data can be assessed only {\it after} a procedure for extracting {\it a continuum spacetime manifold and geometry} out of them has been identified. This is because it is the last notions only that define what space and time are, in our current understanding. We do not use any discrete or algebraic notion of space and time to conceptualize other experiences. The fact that specific discrete quantum structures can be motivated from continuum spacetime theories (or used to approximate them, as in classical Regge calculus) is important and gives us an argument for trusting them as the potentially correct {\it seeds} for the spacetime to come, but not much more. 
For example, the graph or cellular structures  used in LQG, GFT, and simplicial gravity, define a notion of locality that can be interpreted as encoding {\it correlations} between \lq\lq spacetime building blocks\rq\rq, but may differ drastically from any notion of locality emerging in some approximation alongside a continuum spacetime and geometry. Consequently, one should be aware of the risks of the interpretation of lattices as \lq\lq discrete spacetimes\rq\rq {\it over which things happen or move}. This view could be too anchored to the usual background dependent context of lattice gauge theory, where the lattice is indeed a sample of possible paths on an existing background (flat) spacetime. 

The conclusion is that, we argue,  we should interpret the discrete structures suggested by quantum gravity approaches {\it realistically}, i.e. with some trust, but probably {\it not literally}, i.e. with great conceptual care.

\section{Emergent spacetime: what can it mean?}
There are two main classes of objections to the idea of {\it emergent spacetime}. Some have to do more generally with the very idea of {\it emergence}, a rather slippery one. Others are directly tied to the possibility that {\it spacetime}, in particular, could be understood as an emergent concept.

One problematic point is whether or not the supposedly emergent property of a system one wants to describe can be {\it reduced} to properties of more fundamental constituents of the same system, and whether or not the theoretical description of such emergent property can be {\it deduced} from the theoretical description of the fundamental constituents. We will ignore the subtle distinctions between the notions of reduction and deduction, and conflate the two, basically in the spirit of Nagel \cite{nagel}. Can this reduction/deduction be done at all for any property that we may call \lq emergent\rq? A positive answer should come with many qualifications, because to really define and understand this connection is highly non-trivial. This is true \cite{battermann} even for theories whose connection is supposed to be well-understood, like between statistical mechanics and thermodynamics, or between molecular dynamics and hydrodynamics. However, a negative answer, we believe, is untenable. There {\it are} known microscopic derivations of macroscopic phenomena, small scale justifications of large scale effects, even if these reductions may not always be complete, rigorous, analytic, or even, sometimes, useful in practice. 
But if the answer is, roughly speaking, positive, what does it mean that something is \lq\lq emergent\rq\rq at all? Doesn't reducibility of properties to one another imply that the one deduction starts from is {\it the real or truly fundamental one}?.  

As for most of the following, we align with Butterfield and collaborators \cite{butterfield} in understanding {\it emergence} to mean the appearance of properties of a system that are {\it novel} with respect to other descriptions of the same system (e.g. at different scales, or in different approximations, or with different macroscopic constraints) and {\it robust} in the sense of being reproducible and stable, thus systematically observed at least in principle). 
The typical example of reduction/deduction and of emergence that we have in mind is, as in \cite{butterfield}, that of collective properties of many degrees of freedom, collective behaviour that on the one hand can understood to be the result of interactions among \lq\lq basic building blocks\rq\rq, and on the other hand is not \lq\lq implicit in or a simple cumulative effect of\rq\rq the same building blocks. Obviously, the richest realm for studying this type of behaviour is condensed matter theory, and it is also where we will take our specific guiding example from.  

So how can emergence and reducibility be reconciled? We see well justified in this context the solution proposed by Butterfield et al \cite{butterfield}, based on stressing the importance of limits and approximations, and of singular ones and divergences in particular \cite{battermann}. Deducing the emergence of some property requires in general a {\it limiting procedure} in terms of some parameter of the system which leads to a new feature that was not present in (the description of) the system before the limit. This, at the same time, does not imply that the property exists {\it only for the system at the limit}, or that {\it the limit system is real}, as in many cases this would be physically untenable. Here enters the second key notion of {\it approximation}: emergent behaviour, even when it is identified or deduced via a limiting procedure, occurs physically {\it before} the actual limit is reached, provided it is {\it approached enough} in some sense that varies from case to case. It is this emergent behaviour \lq\lq close to the limit\rq\rq which is real. The typical example \cite{butterfield} is phase transitions in condensed matter systems and different emergent features in different phases of a physical system, where the limit to be taken is the one of very large numbers of degrees of freedom, i.e. (for finite density) the thermodynamic limit. 

It is in this sense that we could have reduction and novelty at the same time. Whether reduction is useful or necessary in practice, it is a different debate, which we do not go into \cite{emergence}. Last, we would suggest that the existence of emergent properties forces upon us a more flexible notion of ontology, one that assigns reality to several levels of description of physical systems, and in which there is no reductionism in ontology even in presence of reduction/deducibility in the theoretical description.

\

Coming to the specific case of {\it emergence of spacetime}, a new type of objections is brought forward. 
A first issue is that in a background independent context, and even more in a spacetime-free theory, it is even more difficult to make the notion of emergence clear. Given our tentative general definition above, and assuming we have identified the right \lq\lq fundamental\rq\rq degrees of freedom and the right dynamics for them, and that the nature of the emergent property we want to obtain is clear, i.e. continuum spacetime and geometry and their relativistic dynamics, the question becomes: what is the approximation/limiting procedure such that one can go from one to the other and what are the conditions for its existence? if such procedure exists and if the fundamental degrees of freedom are not themselves continuum spacetimes and geometries, we would have these notions \lq\lq emerging\rq\rq along the way (this is also, more or less, the scheme proposed in \cite{H-W}). We will discuss a specific proposal for what this procedure can be, and an example of it, in the following. 

The above issue has to do with the specific formalism one uses, the strategy adopted, and has to be addressed in examples. Here we only make the general remark that in any quantum gravity approach like the ones introduced above, where the fundamental degrees of freedom are discrete {\it and} quantum, based on combinatorial, rather than continuum, structures, and subject to quantum fluctuations (kinematical superposition as well as probabilistic evolution). There are {\it two} very different types of approximations/limits that have to be taken in order to start from them and arrive to classical GR: a {\it continuum} limit and a {\it classical} limit. Not only the precise definition of both of them will have to be specified, but also the order in which they have to be taken is not obvious at all. It is well possible, as we will show, that a continuum spacetime stems from the quantum properties of its fundamental building blocks, and would not be achieved if only their classical properties were considered.

A more general objection against the idea of emergent spacetime has to do with {\it empirical coherence}. It was put forward, for example, by T. Maudlin in \cite{maudlin}, it is central also in \cite{esfeld}, and was, in our opinion, nicely discussed and counter-argued in \cite{H-W}. The basic point is that in all our theories the connection to observation, thus the empirical relevance of the theories themselves, is ensured by the existence, in the theory, of {\it local beables}, that is observables entities localized in spacetime. In absence of such local beables the theory has no physical relevance. A theory without spacetime in its fundamental definition is very difficult to deal with and to interpret in physical terms. Some of the statements by Maudlin amount basically, in our understanding, to emphasizing how radical and conceptually challenging a quantum gravity theory has to be, if it sticks to its promise. They come as a confirmation of the everyday difficulties of  quantum gravity researchers, rather than as a surprise, and can certainly be agreed on. Indeed, another way to put them is simply to re-state the need, for any quantum gravity theory without continuum spacetime and geometry in its fundamental structures, to unravel a procedure by which they emerge in some approximation. In the same approximation the theory would then have to identify non-trivial local beables, and possibly novel ones, that can be observed, to falsify it. Indeed, it is a tough problem.
Stretching the emphasis on local beables to more than this is, in our opinion (in agreement with \cite{H-W}), over-stretching it. It does not follow as necessity that empirical coherence {\it requires} local beables in the fundamental definition of the theory. Local beables can be removed from fundamental ontology and re-appear as emergent concepts/quantities (e.g. phonons can be characterized in terms of local quantities on the fluid, but the microscopic description of the fluid itself cannot, by definition). Maudlin himself seems to acknowledge this, i.e. that local beables could be appropriate only to some level of reality, as approximations. Would this make the fundamental degrees of freedom described by the theory less real? Maudlin seems to think so, when he questions whether the underlying theory, even though it manages somehow to reproduce local beables in some approximation without having them at the most fundamental level, is \lq\lq physically salient\rq\rq. Huggett and W\"utrich correctly, in our opinion, point out that the issue is what makes a theory \lq\lq physically salient\rq\rq. We do not have the expertise nor the space to discuss this point further. We only point out that this would not be such an unfamiliar situation in physics. The question for any formalism postulating a certain set of degrees of freedom as the microscopic explanation of some macroscopic observable phenomenon, and then succeeding to derive the latter from the former, would be whether we can either obtain a direct phenomenological access to the microscopic entities or at least extract some new macroscopic observable phenomenon that would distinctively follow from specific properties of the same (not directly observable) entities. The strict adherence to the need for local beables for empirical relevance would preclude the first option, but not the second. If the answer is positive, the common physicist's attitude would be to declare the microscopic theory \lq\lq salient\rq\rq. One example is the experimental indirect confirmation of atomic theory via brownian motion before atoms could be directly observed.

We will return to this issue after having offered our proposal for what the emergence of spacetime could look like, and a concrete (tentative) example of its realization.

\section{A concrete, non-spacetime example of emergence}
In order to set the stage for our quantum gravity example of emergence of spacetime, and to clarify the above discussion on emergence in general, we give now one concrete example of it, taken from condensed matter theory: Bose condensation in dilute weakly interacting gases (for details, see \cite{PitStr, Leggett, volovik}).

Assume we have a gas of bosonic atoms, described by a non-relativistic quantum field theory  

\begin{eqnarray}
S(\Psi)= \int dtdx \left[\Psi^*\partial_t
\Psi-\Psi\partial_t \Psi^*\,+\,\Psi^\dagger({\bf
x})\left(-\frac{\hbar^2}{2m}\nabla^2 
\right) \Psi({\bf x})\right] 
+{1\over 2}\int dt dx dy ~\Psi^\dagger({\bf x})
 \Psi^\dagger({\bf y})U({\bf x}-{\bf y}) \Psi({\bf y}) \Psi({\bf x}) \quad,
\label{TheoryOfEverything}
\end{eqnarray}

This microscopic (continuum) field theory, in the regime of small fluctuations around the Fock vacuum $|Fock\rangle$ (the \lq\lq no atom\rq\rq state), i.e. when only a few atoms are present, and atom creation/annihilation processes can be neglected (as in our laboratories), gives an entirely satisfactory description in terms of a discrete system: a finite number of atoms, possibly (e.g. if the system is confined in a box) labeled by discrete quantum numbers (their momenta and energies). 
What does happen when a large number (\lq\lq close to infinite\rq\rq) of the same atoms is considered, together with their interactions, in a thermodynamic limit/approximation (infinite volume), that is, a continuum limit? How do we extract an effective dynamics appropriate for such limit? The question does not have a unique answer, because it depends on the macroscopic conditions imposed on the system of atoms, e.g. temperature, pressure, constraints on the density, values of the coupling constants (whether the system is strongly or weakly interacting). In other words, the system may organize itself in different phases. In any case, we need to {\it change vacuum}. For example, we may know from the solution for the free theory (which can be solved exactly), from experiments or by some intuitive understanding of the physics of bosonic particles, that at very low temperature (and given pressure, weak interaction, etc) the atoms will tend to occupy the same quantum state, they will {\it condense}. That is, we may assume that the new relevant vacuum state around which the continuum dynamics will take place is not the Fock vacuum by a {\it condensate state}:

\begin{equation}
|\varphi_0\rangle = e^{\int \varphi_0(x) \hat{\psi}^\dagger(x)}|Fock\rangle = \sum_N \frac{1}{N!} \int dx_1 ...\int dx_N \varphi_0(x_1) ... \varphi_0(x_N) | x_1, ..., x_N\rangle 
\label{condensate}
\end{equation} 

Clearly, this is a guess and it is {\it not} exactly correct, for an interacting system: it is a state in which we have neglected all correlations between atoms and in which all atoms behave exactly the same, and thus have the same wave function $\varphi_0$. Still, we have reasons to believe that it is not too far from the real one (e.g. because the interaction is weak), so we go ahead and extract the dynamics of the system in terms of an equation for the {\it collective variable} $\varphi_0$. It is a collective variable because it describes at once the behaviour of the {\it infinite} atoms of the system (under the assumption that they have more or less condensed). 

From the quantum microscopic dynamics of the system, encoded in the partition function corresponding to \ref{TheoryOfEverything}, one extracts then an effective equation for the function $\varphi_0(x)$. In the crudest approximation it amounts to 1) assuming that the field operator for the atoms writes $\hat{\psi} = \varphi_0\, \hat{I} + \hat{\chi}$; 2) inserting this in the quantum equations of motion (n-point correlations etc); 3) neglecting the dynamics of the fluctuations (over the condensate) $\hat{\chi}$. The resulting {\it classical} non-linear equation for $\phi_0$ is (with some more approximations made on the microscopic interactions etc) the Gross-Pitaevski equation:

\begin{equation}
i \hbar \partial_t \varphi_0 (x,t) \,=\,\left( - \frac{\hbar^2\,\nabla^2}{2m} \,+\, |\varphi_0(x,t)|^2\right) \,\varphi_0(x,t)
\label{GP}
\end{equation}
which, we note, still contains a dependence on $\hbar$, and a new effective coupling constant resulting from the approximations made on the microscopic interactions \cite{PitStr, Leggett}. This is an hydrodynamic equation for the whole fluid that the condensed atoms form. Its hydrodynamic character can be made explicit by splitting the complex function $\varphi_0$ into its modulus $\rho(x,t)^{1/2}$, corresponding to the fluid density, and its phase, giving the velocity of the fluid as $\vec{v}(x,t) = \vec{\nabla} S(x,t)$, using: $\varphi_0 = \sqrt{\rho} e^{i S}$. It is clear we are dealing with an {\it emergent} behaviour of the whole system of atoms that is not implicit in the individual atoms and crucially depends on having a large number of them and of having taken a continuum (thermodynamic) limit. A new entity has emerged, the fluid, where in the previous description of the same system we had atoms. That it is a robust behaviour is something that can be verified everyday in the laboratories and has to do with collective symmetries (or breaking of them) and their resulting stability, with a degree of universality in microscopic behaviour leading to the same macroscopic one (in similar conditions of temperature, density etc), and with the separation of scales between micro and macro \cite{battermann}. That is is novel, with respect to the atomic description, is another clear fact. It is enough to think of the associated superfluidity.

A few more comments on the issue of emergence in light of this simple example. Even though the macroscopic behaviour of the fluid is to some extent universal and independent on the details of the microscopic system, i.e. the atoms, the microscopic theory is not at all irrelevant. Not only the parameters of the atomic model enter the macroscopic equations, but the type of variables used in the macroscopic theory are closely related to the microscopic ones, even though the dynamics may be vastly different. Moreover, at least in this case there {\it is} a path from the microscopic theory to the macroscopic one and {\it some} macroscopic properties are found to depend on features of the microscopic degrees of freedom, and some features of the atomic dynamics do enter the effective hydrodynamics (this is clear, for example, in spinor BECs). For example, the fact that $\hbar$ appears in the effective classical hydrodynamic equation stems from the specific dispersion relation of the atoms. More generally, the fact that {\it quantum properties} of the atoms are responsible for some of the macroscopic behaviour of the fluid is apparent in the whole phenomenology of superfluidity. 

Now suppose that we did not have access to the atoms at all, experimentally, but still, because we were so smart, we had managed to guess the microscopic theory \ref{TheoryOfEverything}, and then even so smart as to guess the appropriate approximate ground state of the system, in the macroscopic phase we had experimental reasons to believe we lived in, i.e. the state \ref{condensate}. Suppose that we insisted on the idea that the macroscopic fluid we lived in (yes, we are supposing we are unfortunate but smart fishes swimming in the cold superfluid) was only {\it emergent} from something totally different, and unaccessible, and almost unthinkable, the \lq\lq atoms of the fluid\rq\rq. Suppose that we had managed even to find an approximate derivation of \ref{GP} from \ref{TheoryOfEverything}, and thus to hypothetically explain some features of the fluid in terms of its imaginary atoms. In such a situation, would the atomic theory \ref{TheoryOfEverything}, despite not allowing \lq fluid beables\rq to describe the atoms, still be \lq\lq physically salient\rq\rq?

\section{The idea of geometrogenesis}
So, what is the picture of quantum gravity and of the emergence of spacetime that we suggest? The main hypothesis is that what we call continuum spacetime is but a phase of an underlying system of fundamental non-spatio-temporal degrees of freedom, of the type we introduced above, to be reached in a quantum gravity analogue of the thermodynamic limit used in condensed matter systems (for an overview of different quantum gravity models of emergent spacetime, see \cite{lorenzo}). A continuum spacetime would correspond to a collective, emergent configuration of a large number of quantum gravity building blocks. The notion of continuum geometry would, accordingly, make sense only at this emergent level and possibly only in such phase (other continuum phases could be, in principle, non geometric). Emergence of spacetime becomes a problem akin to the emergence of large scale, collective behaviour from atoms in condensed matter theory. In fact, one can be more specific and put forward the hypothesis that spacetime is indeed the result of a {\it condensation} of its microscopic building blocks turning it in a very peculiar type of quantum fluid \cite{GFTfluid}. Actually the idea of spacetime as a condensate has been argued for by several authors \cite{hu, wilczek, laughlin}, and from a variety of standpoints. It can be seen as a way to take seriously the results of analogue gravity models in condensed matter \cite{volovik,bain}, or as the next step in the process of understanding fundamental interactions via spontaneous symmetry breaking (which is the usual particle physics way to formulate the idea of condensation) \cite{wilczek}. It is the extension to spacetime physics of the new (emergent) \lq emergentist\rq\rq paradigm, itself raising from the amazing successes of condensed matter physics in the last century \cite{laughlin}. It is a bit of all of the above, plus the result of a broader view on cosmology, also informed by the various quantum gravity-related arguments for the disappearance of spacetime that we cited at the beginning of this contribution \cite{hu}.

In particular, we have in mind the group field theory framework as the underlying description of spacetime, and its proposed building blocks with features shared with loop quantum gravity, spin foams and simplicial gravity. In this framework, the problem becomes treatable, at least in principle, with more or less standard tools from (quantum) statistical field theory, in particular the renormalization group \cite{GFTrenorm, vincent}, and one can keep a close formal contact with the field theory description of real condensed matter systems. We will show one example in which one takes direct advantage of these formal similarities in the following.

Here we want to comment on the general picture. 

\

First of all, this hypothesis could be seen only as a suggestion for solving the \lq technical\rq problem of deriving a continuum approximation from a candidate fundamental discrete and quantum theory, and a suggestion for the right mathematical tools to use. In this case, however, it would not necessarily carry immediate physical significance, and in particular agreeing with it would not force a realistic interpretation for the building blocks one starts from. 

If one does give physical meaning to the fundamental degrees of freedom suggested by the given approach, then one is compelled to really see the system like a sort of condensed matter system with the General Relativistic  dynamics of spacetime being a sort of hydrodynamic approximation, which would break down if one was able to test the appropriate regime. What this regime can be, exactly, not only is a difficult question given that the very notions of energies and distances are geometry-dependent, but is also something that can be given meaning to only within a specific formalism in which the condensation and the hydrodynamic approximation are realized in the first place. The answer, then, has to wait.

One can however push the realistic interpretation even further, and put forward a further hypothesis. That is, identify the process of quantum spacetime condensation with a known, even if not understood, physical process: the big bang singularity. Better, we identify the coming of the universe, that is of space and time, into being with the physical condensation of the \lq\lq spacetime atoms\rq\rq. There was no space and not time before this condensation happened\footnote{The \lq tensed\rq wording is inevitable and can only refer to some internal time variable, which started running monotonically from the condensation onwards, just like the corresponding tensed statements about the evolution of the universe in (quantum) cosmology; for example, this variable could be an hydrodynamic variable corresponding to the volume of the universe.}.  
Therefore we could call the spacetime condensation {\it geometrogenesis}. In this line of thought \cite{hu, volovik}, cosmological evolution is understood as a relaxation process after the phase transition, towards the equilibrium condensed state. This condensed state would then most likely correspond to some special type of spacetimes, e.g. those described by geometries high degree of symmetries. It is exactly this symmetric geometries, in particular homogeneous ones, that are used in physical cosmology, at late times after the big bang. The tentative example of emergence of spacetime we will present next is too preliminary and recent to confirm this scenario in any compelling way. However, it is consistent with it, as we shall see.

From this perspective, cosmological singularities, i.e. divergences in curvature invariants in GR, could be a sign of the breaking down of the continuum description of spacetime in yet another sense: they would signal the breakdown of the hydrodynamic approximation of the system of spacetime atoms, and at the same time signal the onset of a phase transition for the same system, which we may call, with no disrespect intended, the \lq\lq boiling before the evaporation of the universe\rq\rq. This may even have a more concrete mathematical implementation if, in the same quantum gravity formalism in which the condensation process is described, curvature invariants emerge as macroscopic quantities playing the role of thermodynamic potentials. Then, their divergence would fit with the usual  {\it definition} of a phase transition.

\

If these hypotheses are correct, there is no reason to try to {\it derive} the microscopic dynamics of the quantum building blocks of spacetime, in any strict sense, from continuum GR dynamics. Quantum spacetime dynamics is not quantum GR. The microscopic\footnote{As we remarked concerning the use of \lq tensed\rq sentences, also the use of micro/macro in all this discussion has to be taken in a rather metaphoric sense, and for the same reason (background independence, i.e. no given geometry in the game). We hope the sense of the metaphor is clear from the material presented so far.} dynamics can be some simplified, reduced-to-the-backbone version of GR, as a simplicial version of it may be, but it is not to be expected to be closer than that to the continuum geometrodynamics we know. There are two orders of reasons, both well exemplified in the case of real condensates. The effective dynamics obtained in the hydrodynamic, mean field approximation is  in general very different from the true, microscopic one. Remember, by the way, that we naively expect the quantum gravity description of spacetime to be relevant around the Planck scale, that is around seventeen orders of magnitude (in length) below the quark scale and even more below where we know the GR dynamics to be appropriate. Second, quantizing BEC hydrodynamics, for example, would give operators (observables) corresponding to the density and the velocity of the whole fluid. What happens in that case, though, is the not-so-surprising fact that, before quantum fluctuations of $\hat{\rho}$ or $\hat{\bf v}$ become relevant, the whole hydrodynamic approximation breaks down, and the microscopic atomic structure of the fluid becomes relevant. 

Finally, if the {\it spacetime condensate} and the {\it geometrogenesis} hypotheses are correct, the emergence of spacetime continuum and geometry will be the result of the {\it quantum} properties of the atoms of spacetime. It will be a {\it quantum} phenomenon. Therefore, the order in which the two key limits/approximations needed to recover GR, the semi-classical and the continuum limit, have to be taken is clear: one has first to understand the continuum limit of the quantum system and only then one can take a classical limit and hope to recover a GR-like dynamics. This picture, in the GFT framework, will be substantiated by the forthcoming example.

\section{An example of geometrogenesis}
We now give the anticipated explicit example of emergent spacetime in the context of the GFT framework. We report on some work in progress \cite{GFTcosmology}, along a line of research that was envisaged in \cite{GFTfluid} (see also \cite{lorenzo, danielelorenzo, vincent}), aimed at extracting cosmological dynamics directly from microscopic GFT models, exactly from the idea of continuum spacetime as a condensate, possibly emerging from a big bang phase transition. We do not go into the details of the derivation, also because some such details are being still worked out, and depend on the specific GFT model one uses. We limit ourselves to an outline of the procedure and of the main ideas and results. It should serve the purpose of clarifying many of the general considerations made above, and show a set of concrete possibilities that may have important conceptual implications, and that can be a good basis for further philosophical investigations of the emergent spacetime idea.

\

We have seen that GFTs, just as the field theories describing the fundamental atoms in condensed matter systems, are defined usually in perturbative expansion around the Fock vacuum. In this approximation, they describe the interaction of quantized simplices and spin networks, in terms of spin foam models and simplicial gravity. The true ground state of the system, however, for non-zero couplings and for generic choices of the macroscopic parameters, will not be the Fock vacuum. The interacting system will organize itself around a new, non-trivial state, as we have seen in the case of standard Bose condensates. The relevant ground states (which, due to diffeomorphism invariance, cannot correspond to minima of an energy functional) for different values of the parameters (couplings, etc) will correspond to the different macroscopic, continuum  phases of the theory, with the dynamical transitions from one to the other being indeed phase transitions of the physical system we call spacetime. The fact that the relevant ground state for a proper continuum geometric phase would probably not be the GFT Fock vacuum can be argued also on the basis of the \lq\lq pre-geometric\rq\rq meaning of it: it is a quantum state in which {\it no pregeometric excitations at all are present}, no simplices, no spin networks. It is a {\it no space state}, the absolute void. It can be the full non-perturbative, diffeo-invariant quantum state around which one defines the theory (in fact, it is analogous to the diffeo-invariant vacuum state of loop quantum gravity \cite{LQG}), but it is not where to look for effective continuum physics. Hence the need to {\it change vacuum} and study the effective geometry and dynamics of a different one.
 
\

The first result of \cite{GFTcosmology} is to define an {\it approximation procedure that allows to associate an approximate continuum geometry to the set of data encoded in a generic GFT state}. This applies to GFT models whose group and Lie algebra variables admit an interpretation in terms of discrete geometries, i.e. in which the group chosen is $SO(3,1)$ in the Lorentzian setting or $SO(4)$ in the Riemannian setting, which we focus on here, and additional (simplicity) conditions are imposed, in the model, to reduce generic group and Lie algebra elements to discrete counterparts of a discrete tetrad and a discrete gravity connection \cite{SF, GFT}. 

A generic GFT state with a fixed number $N$ of GFT quanta will be associated to a set of  $4N$ Lie algebra elements: $\{ B_{I(m)}^{A_mB_m}\}$, with $m=1,...,N$ running over the set of tetrahedra/vertices, $I=1,...,4$  indicating the four triangles of each tetrahedron, $(AB)$ indicating Lie algebra components. In turn, the geometricity conditions we mentioned imply that only three elements are independent for each tetrehadron, and are given by: $B_{i(m)}^{AB}={\epsilon_i}^{jk} e_{j(m)}^A e_{k(m)}^B$, with vectors $e^A_{i(m)}\in\bR^{4}$ (for $i=1,2,3,\;m=1,\ldots,N$). Further, one can consider an action of $SO(4)$ on these variables, of the type: $B_{i(m)}\mapsto \left(h_{(m)}\right)^{-1} B_{i(m)} h_{(m)}\,,\quad e_{i(m)}\mapsto e_{i(m)} h_{(m)}$. From the above variables, one can define also the quantities: $g_{ij(m)} = e_{i(m)}^A\,e_{Aj(m)}$, which are invariant under the mentioned group action. 

Next we imagine the tetrahedra being embedded in a spatial 3-manifold $\mathcal{M}$ with a transitive group action $H$\footnote{The embedding is part of the reconstruction procedure chosen, thus of the way one goes about interpreting the data that the theory provides, but it is not part of the definition of the theory itself.}. The embedding is defined by specifying a \lq location\rq of the tetrahedra, i.e. associating for example one of their vertices with a point $x_m$ on the manifold, and three (tangent) vectors $v_{i(m)}$, defining a local frame and specifying the directions of the three edges incident at that vertex. Now we can interpret the vectors $e_{j(m)}^A$ as continuum tetrad vectors integrated over paths in $\mathcal{M}$ corresponding to the edges of the tetrahedron (specified by the $v_{i (m)}$). We have to assume that the paths are sufficiently \lq small\rq, in the metric that we are about to reconstruct, for the approximation to be consistent, and so that the same metric can be approximated by a flat one along the same paths. 

Then, the variables $g_{ij(m)}$ can be used to define the coefficients of continuum metric at a finite number $N$ of points, as: $g_{ij(m)}=g(x_m)({\bf v}_{i(m)},{\bf v}_{j(m)})$ (and similarly for the tetrad vectors), invariant under the above action of the group $SO(4)$. Clearly, this interpretation depends on the choice of embedding. However, for symmetric manifolds, the group action $H$ identifies a unique, canonical choice of embedding: the group $H$ singles out a basis of left-invariant vector fields at each point and thus a preferred embedding in which these are identified with the vectors $v_{i (m)}$. Having made this natural choice, the reconstruction of metric coefficients $g_{ij (m)}$, at a finite number of points,  from the variables associated to a state of $N$ GFT quanta, depends only on the topology of the assumed symmetric manifold $\mathcal{M}$ and on the choice of group action $H$. 

In particular, one can now define an unambiguous notion of {\it homogeneity} for the reconstructed metric. The approximate metric will be {\it homogenous} if it has the same coefficients $g_{ij (m)}$ at any $m$. This captures rigorously the intuitive notion of the metric being the same at every point. Moreover, it implies that the same metric would be also {\it isotropic} if $H=\mathbb{R}^3$ or $H=SU(2)$ and $g_{ij(m)} = a \delta_{ij}$. 

Notice also that we have discussed here only the intrinsic geometry of the spatial manifold $\mathcal{M}$. The same same reconstruction procedure can be applied to the extrinsic geometry, to be extracted from the conjugate data: the group elements labeling the same quantum GFT states. In a quantum context, it goes without saying that it is not possible to specify both sets of data exactly, due to the uncertainly principle. 

Once more, this will give an {\it approximation} to a continuum metric, in that it allows only to specify a continuum metric at a finite number of points $N$. The number of points can be understood as an approximation scale, as the number of observations in our sampling of a continuum metric. This in turn implies that a GFT state that aims at representing a continuum geometry, i.e. something that can be specified at an infinite number of points and that can be perturbed at each of them independently, should involve a number of quanta $N$ that is on the one end allowed to vary and on the other hand allowed to go to infinity.

\

The second result of \cite{GFTcosmology} is the identification of quantum GFT states that, using the above procedure, can be interpreted as {\it continuum homogeneous quantum geometries}. In fact, in such second quantized setting, the definition of states 
involving varying and even infinite numbers of discrete degrees of freedom is straightforward, and the field theory formalism is well adapted to dealing with their dynamics. 

The crucial point, from the point of view of the previous discussion on emergent spacetime and on the idea of spacetime as a condensate of quantum pregeometric and not spatio-temporal building blocks is that {\it quantum states corresponding to homogeneous continuum geometries are exactly GFT condensate states}. The hypothesis of spacetime as a condensate, as a quantum fluid, is therefore realized in quite a literal way.

The simplest state of this type (one-particle GFT condensate), well-defined in the Fock space of any GFT model with basic field $\phi$ and vacuum $| 0\rangle$, and for which we assume a bosonic quantum statistics, is:

\begin{equation}
|\varphi_0\rangle \, \equiv\, \exp{\widehat{\varphi_0}}\,| 0 \rangle = \sum_{m =1}^{\infty} \frac{1}{m!}\widehat{\varphi_0}^m | 0 \rangle \qquad \text{with}\quad \widehat{\varphi_0} \equiv \int [d g] \varphi_0(g_1,...,g_4) \hat{\phi}^\dagger(g_1,..,g_4)\, , \quad\varphi_0(g_i) = \varphi_0(h^{-1}g_i h)
\end{equation}

This describes a coherent superposition of quantum states of arbitrary number of GFT quanta, all of them described by the {\it same} distribution $\varphi_0$ of pregeometric variables. The function $\varphi_0$ is a collective variable characterizing such continuum geometry, and indeed it depends only on invariant homogeneous geometric data. These are the relevant geometries to consider, for example, in a cosmological setting. It is a second quantized state characterized by the fact that the mean value of the fundamental quantum operator $\widehat{\phi}$ is non-zero: $\langle \varphi_0 | \widehat{\phi}(g_i) | \varphi_0\rangle = \varphi_0(g_i)$, contrary to what happens in the Fock vacuum. Other condensate states can be constructed, in particular quantum states taking into account also 2-particle correlations and realizing the symmetry under the group action in a more natural way. These are analysed in detail in \cite{GFTcosmology}. They share however the same general interpretation as homogeneous geometries and cosmological spacetime condensates.

One would expect that these condensate states describe the system in one particular macroscopic continuum phase. In particular, they should arise dynamically from the no-space state via a cosmological phase transition, that is a {\it geometrogenesis}. In fact, alongside the extraction of effective dynamics for them, from the microscopic theory, and their detailed physical interpretation in cosmological terms, an important avenue of recent and future developments is the study of GFT phase transitions in rigorous terms, in turn based on the results obtained in the simpler tensor models \cite{tensorReview}. Only such analysis could give solid grounds to the realization of such GFT states, within the theory, and, in perspective, in nature.  

\

The third main result of \cite{GFTcosmology} is the extraction of effective dynamical equations for the condensate directly from the fundamental GFT quantum dynamics. While the details of the effective dynamics depend on the specific model considered, and on the condensate state used, one can identify the general form of the equations and their generic features. We discuss them briefly here, because they are conceptually interesting. 

The generic form of the dynamics for the condensate $\varphi_0$ is, schematically:

\begin{equation}
\left(\mathcal{K}_{eff}\varphi_0\right)(g_i)\, +\, \mu\, \int [dg] \, \varphi_0 \,...\,\varphi_0\,\mathcal{V}_{eff} \,=\, 0 
\end{equation}

where $\mathcal{K}_{eff}$ and $\mathcal{V}_{eff}$ are modified versions of the kinetic and interaction kernels entering the fundamental GFT dynamics, reflecting the approximations needed to interpret $\varphi_0$ as a cosmological condensate, i.e. the approximations needed for the reconstruction procedure outlined above to be consistent. They define a linear term and an effective interaction term involving the convolution of the interaction kernel with a number of functions $\varphi_0$. These convolutions are as non-local as those in the fundamental GFT interaction, in their pairing of field arguments. We have thus a non-linear and non-local, Gross-Pitaevskii-like equation for the spacetime condensate function $\varphi_0$. Notice that it is an equation {\it on superspace} that is on the space of (homogeneous) geometries of a symmetric spacetime, rather than an equation on spacetime itself. In this sense it is the same type of equation that are used in continuum quantum cosmology, with the basic variable being a \lq\lq wave function of the universe\rq\rq. However, the differences with respect to quantum cosmology are crucial. For one, the equation is non-linear and thus it cannot be interpreted as an Hamiltonian constraint equation for a wave function. These non-linearities can be interpreted as taking into account indirectly, that is in the context of an equation for homogeneous quantum geometries, the effects of inhomogeneities, resulting from the fundamental interactions of the quantum building blocks of spacetime (again, this is in direct analogy with real atomic condensates). This type of non-linear and non-local equations have been in fact proposed independently and out of purely cosmological considerations in \cite{martin} (see also \cite{GFC}), as a generalization of the usual quantum cosmology setting needed to account for inhomogeneities. Non-linear equations have also been proposed \cite{peter} as a way to solve conceptual issues related to quantum mechanical measurements in quantum cosmology. We find it remarkable that here they arise naturally out of the fundamental definition of the quantum gravity theory.  

Next, this equation arises as a \lq\lq hydrodynamic equation\rq\rq for the spacetime fluid, not as the quantization of a classical symmetry-reduced dynamics. In particular, it could be written (as the Gross-Pitaevskii equation) using the decomposition $\varphi_0(g_i) = |\varphi_0(g_i)| e^{i S(g_i)}$, that is in terms of a probability density on the space of geometries and a quantum phase. While it is directly related to a quantum cosmology equation, it seems to us that one has to be careful in bringing in the usual interpretation of quantum wavefunctions. 

Last, notice that, despite its hydrodynamic character and once more in analogy with the dynamics of usual quantum fluids, no classical limit has been taken to derive it, but only a  continuum limit. In fact, the effective equation is meaningful only because of the quantum properties of the microscopic degrees of freedom, in particular their statistics which allows for condensation to take place, and carries the signature of these underlying quantum properties. In particular, as the GP equation, it depends on $\hbar$.  

\

Further approximations are needed to deal wit the effective equations above (and to solve them). 

For example, one can identify regimes (e.g. the regime of very small coupling constant $\mu$, or where special forms of $\varphi_0$ can be used) in which one gets an effective linear equation, in terms of some new kinetic operator $\mathcal{K}_{eff}^{2}$, which will in general involve derivatives on the group (or Lie algebra) manifold. The effective linear equation would then be of the same type of the Hamiltonian constraint equation used in (loop) quantum cosmology, with   $\mathcal{K}_{eff}^{2}$ playing the role of Hamiltonian constraint operator for homogeneous (but anisotropic) cosmologies. One should expect even this effective linear equation, and not only the general non-linear hydrodynamic equation, to imply corrections to the GR evolution of such cosmological geometries, not only because of quantum corrections but also in the sense of a modified classical dynamics. This is more than welcome. It may imply falsifiability of the theory in light of cosmological observations and possibly new insights into the microscopic origin of macroscopic cosmological phenomena.   

One further approximation would then be a semi-classical WKB approximation of the collective variable $\varphi_0 = |\varphi_0| e^{iS}$ in which the quantum phase $S$ is assumed to vary much more rapidly than the density $|\varphi_0|$. Then the phase can be interpreted as a classical action (again, for homogeneous geometries) and the effective dynamical equation $\mathcal{K}_{eff}^{2}\varphi_0$ becomes a Hamilton-Jacobi-like equation for it. 

Last, one can reduce consideration to the isotropic sector, by assuming that $\varphi_0$ depends only on isotropic geometric variables, that is the overall scale factor $a$ of the universe and its conjugate variable. 

\

To illustrate schematically\footnote{This is supposed to be only a tentative description of what {\it could} happen, since the analysis of the GFT condensate equations is still in progress \cite{GFTcosmology}.} the above approximations, let us consider the simple case in which $\mathcal{K}_{eff}^{2} = \sum_I\Delta_I$, where $\Delta_I$ is the Laplace-Beltrami operator on $SU(2)$ and we assume that the function $\varphi_0$ depends on four such $SU(2)$ variables. This is a realistic example, given that most GFT models for 4d gravity involve geometricity conditions that effectively reduce the GFT field to depend only on $SU(2)$ variables, and that Laplacian kinetic terms seem to be required for a proper renormalization group analysis and for renormalizability of GFT models. 

With this choice of $\mathcal{K}_{eff}^{2}$, one can introduce coordinates on $SU(2)$ given by $g = \sqrt{1-\vec{\pi}^2}\,{\bf 1} - i \vec{\sigma}\cdot\vec\pi\,,\quad|\vec{\pi}|\leq 1$ and obtain, in a WKB approximation in which $|\varphi_0|$ is considered constant, the equation: $\sum_I\left(B_I\cdot B_I - (\pi_I\cdot B_I)^2\right)=0$, where $B_I:=\partial S/\partial \pi_I$. The last variables can be interpreted as homogeneous triad variables, while the $\pi_I$ correspond to homogeneous connection variables. They can be expressed in terms of cosmological scale factors as: $B_I=a_I^2\,T_I$ and $\pi_I = \dot{a}_I V_I$, where $T_I, V_J$ are (state-dependent) normalized Lie algebra elements. Then the classical dynamical equation reduces to: $\sum_I a_I^4\left(\dot{a}_I^2\, T_I\cdot V_I\,-\, 1\right) = 0$. Last, one can restrict the variables to the isotropic ones by setting $a_I=a$. The final equation one gets (for non-degenerate geometries) is:

\begin{equation}
\dot{a}^2 - k = 0
\end{equation}
that is the Friedmann equation for a homogeneous universe with constant curvature $k = \frac{1}{4}\sum_I T_I \cdot V_I$.

\section{Concluding comments: emergence spacetime, what does it mean, then?}
We conclude by reconsidering the idea of emergent spacetime and in particular the idea of spacetime as a quantum condensate., in light of the GFT example we have just discussed.  

The above example represents the first derivation of cosmological continuum spacetime and geometry, including its dynamical aspects, from a microscopic quantum gravity theory, as far as we are aware of. This is remarkable enough, in our opinion. However, it should be noted that it may represent a solution to the problem of the emergence of continuum spacetime and geometry in quantum gravity only in the simplest case in which truly local degrees of freedom are absent, which is the case for homogeneous geometries. Alongside the description of the GFT condensation as a phase transition, thus realizing the idea of geometrogenesis, the study of cosmological perturbations is indeed the next step in the research programme centered around the idea of GFT condensation, and can be tackled as well using ideas from condensed matter theory. 

Homogeneous geometries represent a very coarse grained level of description of spacetime, and a detailed implementation of such coarse graining is very difficult, bot at the classical \cite{homCosmo} and quantum level \cite{bianca, coherent2}. The mechanism of condensation bypasses many such difficulties and allows to related directly the microscopic quantum dynamics to the macroscopic coarse grained one. Still, it is clear that the real outstanding issue is to describe the \lq middle ground\rq, i.e. the intermediate regime between the non-spatio-temporal quantum structures of the fundamental theory and the simple coarse grained structures at cosmological scale, i.e. the regime that one could expect to be described by some modified version of inhomogeneous quantum GR.

\

Still, the example clarifies what the realization of the emergent spacetime idea may entail. We see realized in it several features of the general scheme suggested by Huggett and W\"utrich \cite{H-W}: quantities with a genuine continuum geometric interpretation are recovered by means of an approximation procedure; the superposition of quantum states plays a central role in this recovery of continuum structures, and at the same time the emergent spacetime does not immediately encode adjacency relations (microscopic locality) characterizing the microscopic discrete quantum structures. 

At a more general level, the example shows the important role of limiting procedures in terms of parameters controlling the number of degrees of freedom involved from the microscopic theory, here the number of GFT quanta $m$, as advocated by Butterfield and collaborators \cite{butterfield}. And this limiting procedure may in turn lead to the emergence of continuum spacetime being signaled by divergences, singularities in macroscopic quantities, as advocated in \cite{battermann}, as it would be the case if the GFT condensate is realized after a phase transition of the relevant GFT model. 

At the same time, and again in accordance with \cite{butterfield}, the parameter $m$ encodes the approximation scale, the level of detail of a sampling of continuum geometries by GFT data, and one could well imagine a finite but large enough $m$ to capture the essential features of macroscopic cosmological dynamics, in such a way that one does not have to assume that the \lq\lq infinity\rq\rq of GFT quanta is physically realized in nature.

Last, it shows that the fear of empirical incoherence may not be justified. If the details of the above procedure can be worked out nicely, and once the same procedure is generalized to deal with inhomogeneities, we would have a concrete framework for emergent spacetime in which local, continuum spacetime properties are non \lq fundamental\rq, that is not present in the non-spatio-temporal quantum degrees of freedom that \lq constitute\rq\rq spacetime, and out of whose condensation the latter emerges, but at the same time become real, empirically relevant quantities in some controlled approximation, in one specific regime of the fundamental theory. In this regime, the effective dynamics would carry the signature of microscopic quantum structures which would be, in turn, empirically relevant because indirectly falsifiable via their macroscopic consequences.

In this last respect, the GFT example, just as the BEC example, confirms the compatibility between genuine emergence and reduction, as argued in \cite{butterfield}. Continuum spacetime geometry can be the result of {\it quantum collective behaviour} of microscopic non-spatio-temporal degrees of freedom. It can be a novel and  robust collective, thus {\it emergent},  description of them, but at the same time it can be deduced from the microscopic non-spatio-temporal description, in such a way that the origin of macroscopic phenomena can be traced back, in principle, to specific microscopic properties, even if not necessarily derived from them analytically or rigorously in all detail. 

\

The picture of spacetime as a condensate, emerging from a non-spatio-temporal and non geometric phase through geometrogenesis, challenges profoundly our worldview, in its most fundamental aspects. It forces a rethinking of our basic ontology, in two ways. Its \lq emergentist\rq approach applied to the most basic structures of the world, space and time themselves, makes any simple-minded reductionist or fundamentalist ontology problematic. If spacetime emerges in a concrete, testable setting, from non-spatio-temporal structures from which it can be, however, precisely derived, then it is not {\it real} in an ontological sense, if such reductionist ontology is maintained. And this conclusion is usually anathema for the same thinkers who maintain reductionist ontologies in the first place. If one, on the other hand, wants to maintain that space and time are real, that they do exist, then one seems to be forced toward a more flexible ontology, in which different levels of reality, different regimes of approximation and different modes of organization of physical systems in the world, are all equally existent, all have a fundamental ontological status. And the fact that one can reduce one to another, or deduce one from the other, by whatever procedure, is in itself no argument for the existence of one and the non-existence of the other.  Liquid water is just as real, in such fundamental ontology, as the molecules of hydrogen and oxygen that constitute it, and that we experience in different regimes of the same system and in different approximations of its description, and just as ontologically real as solid ice or vapor. This may sound obvious to some, but the point here is that the realization of the \lq spacetime as condensate\rq idea may force us to adopt the same flexible ontology for space and time themselves. This will be first done as a speculation, as a working hypothesis, but this new attitude would then suggest to pose new concrete questions to the framework realizing it: can we access experimentally, directly that is, the \lq\lq atoms of spacetime\rq\rq? can we identify physical situations in which the \lq\lq other phases\rq\rq of the same system, the ones that do not correspond to continuum, geometric spacetime, are realized in nature?

Ontological questions would become scientific questions, at an even more fundamental level than the analogous questions regarding material atoms more than a century ago.

\section*{Acknowledgements}
This work has been supported by the Alexander von Humboldt Stiftung, through a Sofja Kovalevskaja Prize, which is gratefully acknowledged. We would also like to thank all the participants to the workshop \lq Reflections on space, time, and their quantum nature\rq, held at the Albert Einstein Institute in November 2012, for many interesting discussions on these matters.


\begin{thebibliography}{99}
\bibitem{H-W}  N. Huggett, C. W\"utrich, to appear in Studies in the History and Philosophy of Modern Physics, arXiv:1206.6290v2 [physics.hist-ph]
\bibitem{LQG} T. Thiemann, \textit{Modern canonical quantum General Relativity}, Cambridge University Press, Cambridge (2007); C. Rovelli, {\it Quantum Gravity}, Cambridge University Press, Cambridge (2006); C. Rovelli, PoS QGQGS2011 (2011) 003, arXiv:1102.3660 [gr-qc]
\bibitem{carlo} C. Rovelli, Found.Phys. 41 (2011) 1475-1490, arXiv:0903.3832 [gr-qc]
\bibitem{LQC} A. Ashtekar, P. Singh, Class.Quant.Grav. 28 (2011) 213001, arXiv:1108.0893 [gr-qc]
\bibitem{time} C. Isham, in Salamanca 1992, Proceedings, Integrable systems, quantum groups, and quantum field theories* 157-287, gr-qc/9210011
\bibitem{SF} A. Perez, to appear in Living Reviews, arXiv:1205.2019 [gr-qc] 
\bibitem{doplicher} S. Dolplicher, K. Fredenhagen, J. Roberts, Commun.Math.Phys. 172 (1995) 187-220, hep-th/0303037 
\bibitem{sabine} S. Hossenfelder, to appear in Living Review on Relativity,  arXiv:1203.6191 [gr-qc]
\bibitem{RelatLoc} G. Amelino-Camelia, L. Freidel, J. Kowalski-Glikman, L. Smolin, Gen.Rel.Grav. 43 (2011) 2547-2553, Int.J.Mod.Phys. D20 (2011) 2867-2873, arXiv:1106.0313 [hep-th]
\bibitem{DSR} J. Kowalski-Glikman, in {\it Approaches to Quantum Gravity}, D. Oriti (ed.), Cambridge University Press, Cambridge (2009), gr-qc/060302
\bibitem{strings} S. Giddings, Phys.Rev. D74 (2006) 106006, hep-th/0604072 
\bibitem{stringdualities} D. Rickles, Stud.Hist.Philos.Mod.Phys. 42 (2011) 54-67; G. Horowitz, J. Polchinski, in {\it Approaches to Quantum Gravity}, D. Oriti (ed.), Cambridge University Press, Cambridge (2009),  gr-qc/0602037
\bibitem{seiberg} N. Seiberg,  hep-th/0601234, T. Banks, Int.J.Mod.Phys. D21 (2012) 1241004
\bibitem{nicolai} T. Damour, H. Nicolai, Int.J.Mod.Phys. D17 (2008) 525-531,  arXiv:0705.2643 [hep-th]
\bibitem{giddings} S. Giddings, Class.Quant.Grav. 28 (2011) 025002, arXiv:0911.3395 [hep-th]  
\bibitem{tedrenaud} T. Jacobson, R. Parentani, Found.Phys. 33 (2003) 323-348, gr-qc/0302099
\bibitem{sorkinBH} R. Sorkin, Stud. Hist. Philos. Mod. Phys. 36 (2005) 291-301,  hep-th/0504037
\bibitem{jacobson} T. Jacobson, Phys.Rev.Lett. 75 (1995) 1260-1263, gr-qc/9504004
\bibitem{analog} C. Barcelo, S. Liberati, M. Visser, Living Rev.Rel. 14 (2011) 3,  gr-qc/0505065 
\bibitem{sindoni} L. Sindoni, Phys.Rev. D83 (2011) 024022, arXiv:1011.4411 [gr-qc]; S. Finazzi, S. Liberati, L. Sindoni, Phys.Rev.Lett. 108 (2012) 071101, arXiv:1103.4841 [gr-qc]
\bibitem{hamber} H. Hamber, arXiv:0704.2895 [hep-th]
\bibitem{CDT} J. Ambjorn, J. Jurkiewicz, R. Loll, Physics Reports 519 (2012) 127-210, arXiv:1203.3591 [hep-th]
\bibitem{mm} P. Di Francesco, P. Ginsparg, J. Zinn-Justin, Phys. Rept. 254 (1995) 1-133, arXiv: hep-th/9306153
\bibitem{CS} F. Dowker,  gr-qc/0508109; D. Benincasa, F. Dowker, Phys.Rev.Lett. 104 (2010) 181301, arXiv:1001.2725 [gr-qc]
\bibitem{causalSF} E. Livine, D. Oriti, Nucl.Phys. B663 (2003) 231-279, gr-qc/0210064
\bibitem{GFT} D. Oriti,  in G.~Ellis et al. (eds.), \textit{Foundations of space and time}, Cambridge University Press, Cambridge (2012), arXiv:1110.5606 [hep-th]; D. Oriti, in: D. Oriti (ed.), \textit{Approaches to quantum gravity},  Cambridge University Press, Cambridge  (2009), arXiv:  gr-qc/0607032; D. Oriti, in {\sl Quantum gravity}, Fauser, B. (ed.) et al., 101-126, Birkhauser (2006), gr-qc/0512103 [gr-qc]; D. Oriti, arXiv:0912.2441 [hep-th]
\bibitem{aristidedaniele} A. Baratin, D. Oriti, Phys.Rev.Lett. 105 (2010) 221302, arXiv:1002.4723[hep-th]; A. Baratin, D. Oriti, Phys.Rev. D85 (2012) 044003, arXiv:1111.5842 [hep-th]
\bibitem{GFTrenorm} S. Carrozza, D. Oriti, V. Rivasseau, arXiv:1207.6734 [hep-th]; V. Rivasseau, AIP Conf.Proc. 1444 (2011) 18-29, arXiv:1112.5104 [hep-th]; V. Rivasseau, PoS CNCFG2010 (2010) 004, arXiv:1103.1900 [gr-qc];   J.~Ben Geloun,
  arXiv:1205.5513 [hep-th]
\bibitem{vincent} V. Rivasseau,  arXiv:1209.5284 [hep-th]
\bibitem{tensorReview} R. Gurau, J. Ryan, SIGMA 8 (2012) 020, arXiv:1109.4812 [hep-th]
\bibitem{nagel} E. Nagel, {\it The Structure of Science: problems in the Logic of Scientific Explanation}, Harcourt (1961)
\bibitem{battermann} R. Batterman, Found. Physics, 41, 6, 1031-1050 (2011); R. Batterman, in: The Robert and Sarah Boote Conference in Reductionism and Anti- Reductionism in Physics, (Pittsburgh, April 22-23, 2006).
\bibitem{butterfield} J. Butterfield, Foundations of Physics 41, 2011, 920-959; J. Butterfield, Foundations of Physics 41, 2011, 1065-1135; J. Butterfield, N. Bouatta, in Proceedings of Frontiers of Fundamental Physics 11 (AIP), eds. J. Kouneiher, C. Barbachoux and D. Vey, 2012
\bibitem{emergence} M. Bedau, P. Humphreys (eds), {\it Emergence}, MIT Press, Cambridge MA (2008)
\bibitem{maudlin} T. Maudlin, J. Physics, 40, 3151-3171 (2007)
\bibitem{esfeld} M. Esfeld, V. Lam, to appear in Studies in History and Philosophy of Science B   
\bibitem{butterfield2} J. Butterfield, Interface Focus (Royal Society London) 1, 2011, 1-14; G. Ellis, Interface Focus, 2,1 126-140 (2012)
\bibitem{PitStr} L. Pitaevskii, S. Stringari, {\it Bose-Einstein Condensation}, Oxford University Press (2003)
\bibitem{Leggett} A. Leggett, {\it Quantum Liquids}, Oxford University Press (2006); A. Leggett, Rev. Mod. Phys. 73, 307Ð356 (2001)
\bibitem{volovik} G. Volovik, gr-qc/0612134; G. Volovik, Phys.Rept. 351 (2001) 195-348, gr-qc/0005091
\bibitem{GFTfluid} D. Oriti, D. Oriti, Proceedings of Science PoS(QG-Ph)030, [arXiv:0710.3276]; L.~Sindoni, arXiv:1105.5687 [gr-qc]; D. Oriti, J.Phys.Conf.Ser. 67 (2007) 012052, arXiv: hep-th/0612301
\bibitem{hu} B-L. Hu, Int.J.Theor.Phys. 44 (2005) 1785-1806, gr-qc/0503067; B-L. Hu,  gr-qc/9511076
\bibitem{wilczek} F. Wilczek, {\it The Lightness of Being}, Basic Books (2010)
\bibitem{laughlin} R. Laughlin, {\it A Different Universe}, Basic Books (2006)
\bibitem{bain} J. Bain, Studies in History and Philosophy of Modern Physics (2012)
\bibitem{lorenzo} L. Sindoni, SIGMA 8 (2012) 027, arXiv:1110.0686 [gr-qc]
\bibitem{danielelorenzo} D. Oriti, L. Sindoni, New J.Phys. 13 (2011) 025006, arXiv:1010.5149 [gr-qc]
\bibitem{GFTcosmology} S. Gielen, D. Oriti, L. Sindoni, {\it Cosmology from Group Field Theory}, to appear (2013)
\bibitem{martin} M. Bojowald, A. Chinchilli, C. Dantas, M. Jaffe, D. Simpson, Phys.Rev. D86 (2012) 124027, arXiv:1210.8138 [gr-qc]
\bibitem{peter} J. Martin, V. Vennin, P. Peter, Phys.Rev. D86 (2012) 103524, arXiv:1207.2086 [hep-th]
\bibitem{GFC} G. Calcagni, S. Gielen, D. Oriti, Class.Quant.Grav. 29 (2012) 105005, arXiv:1201.4151 [gr-qc]
\bibitem{homCosmo} T. Buchert, Class.Quant.Grav. 28 (2011) 164007, arXiv:1103.2016 [gr-qc]
\bibitem{bianca} B. Dittrich, F. Eckert, M. Martin-Benito, New J.Phys. 14 (2012) 035008, arXiv:1109.4927 [gr-qc]
\bibitem{coherent2} D. Oriti, R. Pereira, L. Sindoni, Class.Quant.Grav. 29 (2012) 135002, arXiv:1202.0526 [gr-qc]

\end{thebibliography}
\end{document}